# On a Method for Multiscale Solid Mechanics

Amit Acharya

Dept. of Civil and Environmental Engineering

Carnegie Mellon University, Pittsburgh, PA 15213, U.S.A.*

**Abstract**

Given a fine-scale physical theory characterized by an evolutionary system of equations and a set of quantities, defined from the variables of the fine theory, that serve as a coarse representation of the fine scale phenomena, a systematic method for developing an evolutionary set of equations for the coarse quantities is presented. This set may be considered as governing equations for a coarse theory corresponding to the fine one since the solutions of the coarse theory define a small class of solutions to the fine theory. The method for generating the coarse theory is technically straightforward. The conditions required for the procedure to succeed are understood qualitatively, but a rigorous result, for the intended applications, is not known – hence the method is formal. The method is illustrated on a few model problems.

## *1. Introduction*

This paper is concerned with the development of a multiscale method in solid mechanics for nonlinear evolutionary systems of equations. The method reduces to spatial homogenization in most circumstances. Despite the drastic improvements in computational speed and capability, there remain important applications in solid mechanics where coarse scale behavior cannot be computed, due to computational limitations, from known fine scale physical principles. Moreover, even in cases where this may be possible, such an approach may not be warranted depending upon the nature of information required about the system under study. There are many applications in solid mechanics that can benefit from a general homogenization technique for nonlinear evolutionary equations that lends itself to computational determination, so that approximate coarse theories may be generated routinely. Some such examples are averaged single crystal plasticity response based on dislocation mechanics, macroscopic polycrystal plasticity based on single crystal plasticity response of constituent crystal aggregates, average response of porous media, wave propagation of average fields in inhomogeneous elastic media (rocks and soils), averaged response of granular media, and the development of exact elastic and inelastic shell and rod theories from the corresponding 3-d continuum theory.

* Tel. (412) 268 4566; Fax. (412) 268 7813; email: amita@andrew.cmu.edu

To the best of the author's knowledge, there does not exist a general theoretical procedure in the literature for the homogenization of nonlinear evolutionary equations. One approach valid for materially inhomogeneous linear elastodynamics is that of Willis (1981), Sabina *et al.* (1993), Smyshlyaev *et al.* (1993; a,b). A different approach, also valid for the mathematically rigorous study of limits of sequences of solutions to nonlinear hyperbolic conservation laws, is based on the work of L. Tartar, an exposition of which can be found in Tartar (1990). Applications of this idea may be found in McLaughlin *et al.,* (1985), DiPerna (1988), E and Kohn (1991), and Amirat *et al.* (1992), amongst many other works.

In this paper, a procedure different from the two above will be pursued. When the conditions for its applicability are satisfied by a fine scale system under consideration, it appears to be a powerful and technically straightforward tool for determining the structure of a corresponding coarse scale theory.

The method to be pursued is based on Muncaster's (1983; a) Theory of Invariant Manifolds in Mechanics. This procedure has a compelling history in that it was developed as a generalization of the method that delivers continuum hydrodynamics (coarse theory complete with constitutive relations) from a complete knowledge of Maxwell's second kinetic theory of a monatomic gas (fine scale theory), as shown in Truesdell and Muncaster (1980). It has also been applied to the study of pseudo-rigid bodies by Muncaster (1983; b) and Cohen and Muncaster (1984). A study of the analytical problems that are associated with the applicability of the method is presented in Carr and Muncaster (1983; a,b) in simper settings than would arise in the typical solid mechanics application.

Essentially, the method has four requirements to be applicable:
1. A fine scale evolutionary theory.
2. A formula associating to each state of the fine scale theory, a unique coarse state.
3. The existence of an *invariant 'manifold'* (or surface), in the configuration space of states of the fine theory. This manifold has the property that states corresponding to an evolutionary solution related to initial data specified on the manifold stay on the manifold itself.
4. The set of states of the coarse theory should parametrize the above manifold.

When these requirements are satisfied, the products of the multiscale procedure are:
1. A completely defined evolutionary coarse scale theory.

2. A mapping that transforms any solution of the coarse scale theory directly into solutions of the fine theory. This mapping is an integral part of the definition of the coarse theory.

The mapping provides justification for the use of the coarse theory in the sense that any solution of it corresponds, through the mapping, to a solution of the fine theory. Thus, the study of solutions of the coarse theory as a reflection of gross physics is justified, as this also amounts to the study of a restricted class of solutions of the fine scale physics.

The plan of the paper is as follows: in Section 2 we present the theory and some comments on the basis of a possible computational procedure to approximate the equations of the theory. In Section 3, 4, and 5 we apply the theory to three model problems; spatial homogenization of a 1-D, linearly rate-dependent elastoplastic problem; the reduction of degrees of freedom in field dislocation mechanics to obtain a slip gradient based plasticity model; and the structure of homogenized 1-D elastodynamics, provided a solution to the derived equations defining the homogenization procedure exists. We end with some observations on the range of validity of the coarse theory in Section 6.

The motivation for this study arises from a desire to homogenize field dislocation mechanics (Acharya, 2001; a,b) to obtain a theory of macroscopic single crystal plasticity where the description of work hardening, plastic deformation evolution, and stress are outcomes of the homogenization procedure. The complications in achieving this goal are significant. For example, while the dislocation density tensor is the primary state variable in field dislocation mechanics, it is well appreciated that it cannot be an appropriate state variable for a theory at coarser scales of resolution larger than the average dislocation spacing, because of the tendency of dislocation segments to be present in equal numbers with opposite sign in any macroscopic region. However, the volume average of the norm of the dislocation density tensor over specified length scales is certainly non-vanishing, and in cases that the homogenization procedure in question succeeds, a completely defined evolution equation for the volume average of the norm becomes a natural outcome of the theory. It is to be noted that the latter measure of dislocation density is adopted as the primary state variable in macroscopic theories of plasticity and the specification of its evolution is of utmost importance in the prediction of gross behavior of real materials. Much is known empirically (Kocks *et al.*, 1998) about such evolution, and an eventual target of the type of theory being attempted here will be to recover results of the empirically deduced formulations and, of course, to predict additional details of macroscopic material behavior.

The simple model problems solved in this paper at least point to the fact that invariant manifolds on which physically significant macroscopic averages execute their dynamics may not be very elusive, and actually do exist for a class of theories in solid mechanics that may be deemed as fine scale theories. In such cases, coarse scale dynamical theories corresponding to the fine scale ones may be approximated by a fairly straightforward, albeit tedious, procedure. Of course, it would be nice to have a general mathematical result that precisely states and answers the following question: to what extent does the physical expectation of the existence of coarse descriptions corresponding to fine scale phenomena necessarily translate to the existence of invariant manifolds in the phase space of a dynamical description of the fine scale phenomena?

## 2. *The mathematical scheme*

As an application of Muncaster's (1983; a) *formal theory of invariant manifolds*, we consider a restricted framework that is still likely to be applicable to the study of a large class of theories of solid mechanics. Let a *fine* theory be defined by a system of evolutionary equations of the form

$$\dot{f}_I(\bm{x},t) = F_I(f(t))(\bm{x}) \quad \bm{x} \in \bm{B}, t \in \bm{I}, \quad 1 \leq I \leq N$$
$$f_I(\bm{x},0) = f_{0I}(\bm{x})$$

(1)

As for notation in the above and what follows, a superposed dot represents partial differentiation with respect to time. If a letter representing an array appears without indices then it represents the entire collection – this device will be used typically to represent arguments of functions. The summation convention over the range of indices is always invoked, unless the summation rule for the index is explicitly controlled by a summation $\left(\sum\right)$ symbol. Also, for a function $g$ of space and time, the notation $g(t)$ represents the corresponding function of space at time $t$. In (1) above, $F_I$ represents a functional taking as argument $N$ functions on $\bm{B}$ and delivering a function on $\bm{B}$, e.g. in (1), $f(t)$ is the argument list and $F_I(f(t))$ is the output. For instance, $F_I$ could be a nonlinear differential or integral operator. The set of functions $f_0$ are prescribed, and correspond to initial conditions. Explicitly time-independent boundary conditions may be included in the specification (1), but in general we ignore boundary conditions for the fine theory for the purpose of developing this multiscale method. Clearly, then, there may be non-uniqueness in solutions of (1) but this is not relevant to the subsequent development of the method.

Let us now consider a set of known functions $p_i, 1 \leq i \leq n$, each of which is generally a function of all the fine fields and their first partial derivatives with respect to the space variables. The argument list of the functions $p$ are indexed as $(\cdots, I, \cdots; \cdots, I\mu_1, \cdots)$, $1 \leq I \leq N$, $1 \leq \mu_1 \leq 3$. Given any set of fine fields $f$ (as a function of the space variables), we use the notation

$$p_i(\nabla^1 f(\boldsymbol{x})) = p_i(\cdots, f_I(\boldsymbol{x}), \cdots; \cdots, f_{I,\mu_1}(\boldsymbol{x}), \cdots), \tag{2}$$

and we use this notation henceforth for any function with an argument list identical to that of the $p$s. A set of coarse fields are now defined in terms of the fine fields through the prescription

$$c_i(\boldsymbol{x}, t) = \int_B w_{ij}(\boldsymbol{x} - \boldsymbol{x}') p_j(\nabla^1 f(t)(\boldsymbol{x}')) d\boldsymbol{x}' \quad 1 \leq i \leq n, 1 \leq j \leq n. \tag{3}$$

The weighting functions $w$ are assumed to be known and vanish identically if $i \neq j$. In the intended applications, the weighting functions, when suitably normalized, provide for a precise definition of 'running' weighted averages of physically relevant functions of the fine scale field quantities over specified length scales.

The crucial idea of the formal method of invariant manifolds is the realization that a collection of states of the fine theory representing the invariant manifold may be parametrized by the states of the coarse theory through a functional termed as a *Gross Determiner*, i.e.

$$f_I(\boldsymbol{x}, t) = G_I(c(t))(\boldsymbol{x}), \tag{4}$$

and the main task of the theory becomes the determination of the functionals $G$. In the present context, we restrict the generality of $G$ in (4) in the sense of retaining only select terms from a formal Taylor-type expansion, and then provide a formal procedure for the determination of this restricted functional. If the full generality of (4) is retained, then the procedure for determining $G$ demands the solution of equations involving Frechet derivatives of functionals on their function space domains, as shown in Muncaster (1983; a).

In order to proceed, for the moment consider $G_I$ to be a function of the values of $c(t)$ at a finite number of points in the body. Then

$$G_I(x_1, \cdots, x_\mu, \cdots, c_1(\boldsymbol{x}_1, t), \cdots, c_j(\boldsymbol{x}_A, t), \cdots) = G_I(\boldsymbol{x}, \boldsymbol{0}) + \sum_{j,A} \partial_{jA} G_I(\boldsymbol{x}, \boldsymbol{0}) c_j(\boldsymbol{x}_A, t)$$
$$+ \sum_{j,k,A,B} \frac{1}{2} \partial^2_{jA,kB} G_I(\boldsymbol{x}, \boldsymbol{0}) c_j(\boldsymbol{x}_A, t) c_k(\boldsymbol{x}_B, t) + \cdots. \tag{5}$$

where the symbols for the partial derivative functions simply represent the partial derivatives of the function in question with respect to a corresponding argument in its argument list. Noticing this pattern corresponding to the finite number of dependencies, we assume $G_I$ in (4) to be of the form

$$G_I(c(t))(\bm{x}) = G_{0I}(\bm{x}) + \int_B G_{1Ij}(\bm{x}, \bm{x}') c_j(\bm{x}', t) d\bm{x}' \qquad (6)$$
$$+ \int_B \int_B G_{2Ijk}(\bm{x}, \bm{x}', \bm{x}'') c_j(\bm{x}', t) c_k(\bm{x}'', t) d\bm{x}' d\bm{x}'' + \cdots$$

A corresponding gradient approximation may be deduced from (6) as

$$G_I(c(t))(\bm{x}) = G_{0I}(\bm{x}) + \tilde{G}_{1Ij}(\bm{x}) c_j(\bm{x}, t) + \tilde{G}_{1Ij\mu}(\bm{x}) \partial_\mu c_j(\bm{x}, t) + \cdots \qquad (7)$$
$$+ \tilde{G}_{2Ijk}(\bm{x}) c_j(\bm{x}, t) c_k(\bm{x}, t) + \tilde{G}_{2Ijk\mu}(\bm{x}) c_j(\bm{x}, t) \partial_\mu c_k(\bm{x}, t) + \cdots.$$

Keeping in mind (6) and (7) along with the strong local nonlinearities that occur in empirically developed constitutive theories for inelastic macroscopic response of crystalline materials, whereas the underlying dislocation mechanics at a fine scale, while being nonlinear, is not as strongly nonlinear but more strongly nonlocal, we choose to proceed with the following assumed form for the functional $G_I$:

$$G_I(c(t))(\bm{x}) = \tilde{g}_I(\bm{x}) + \int_B g_{Ij}(\bm{x}, \bm{x}') c_j(\bm{x}', t) d\bm{x}' + \sum_{A=1}^{Y} l_{IA}(\bm{x}) h_A(\nabla^M c(t)(\bm{x})) \qquad (8)$$

where $Y$ and $M$ are arbitrarily chosen positive integers. The values of $Y$ and $M$ may be sequentially increased to improve the nature of the approximation. The functions $h$ are assumed to be known (e.g. polynomials), with a list of arguments indexed as $(\cdots, i, \cdots; \cdots, i\mu_1, \cdots; \cdots; \cdots, i\mu_1 \ldots \mu_\rho, \cdots; \cdots; \cdots, i\mu_1 \ldots \mu_M, \cdots)$, $1 \leq i \leq n$, $1 \leq \mu_\rho \leq 3$, $1 \leq \rho \leq M$. Given any set of coarse fields $c$ (as a function of the space variables), we use the notation

$$h_A(\nabla^M c(\bm{x})) = h_A(\cdots, c_i(\bm{x}), \cdots; \cdots, c_{i,\mu_1}(\bm{x}), \cdots; \cdots; \cdots, c_{i,\mu_1 \ldots \mu_\rho}(\bm{x}), \cdots; \cdots; \cdots, c_{i,\mu_1 \ldots \mu_M}(\bm{x}), \cdots), (9)$$

and we use this notation henceforth for any function with an argument list identical to that of the $h$s. With this chosen form, the objective now is to develop a procedure for determining the functions $\tilde{g}, g, l$. To emphasize the fact that $G$, as represented by (8) is entirely characterized by $\tilde{g}, g, l$, we will often use the notation

$$G_I(d)(\bm{x}) = \Gamma_I(\tilde{g}, g, l, d)(\bm{x}) \qquad (10)$$

for any functions $d$ on $\bm{B}$ (in a suitable class, of course).

To find equations for the determination of the Gross Determiner, let us first assume that there exist functions $\tilde{g}, g, l$ and a solution $f$ to (1) that also satisfy

$$\Gamma_I(\tilde{g}, g, l, c(t))(\bm{x}) = f_I(\bm{x}, t), \qquad (11)$$

with $c(t)$ defined from $f(t)$ through (3) for each $t$ in the time interval $I$ of interest. Then

$$\dot{f}_I(x,t) = \int_B g_{Ij}(x,x')\dot{c}_j(x',t)\,dx' + \sum_{A=1}^{r} l_{IA}(x)\partial_j h_A(\nabla^M c(t)(x))\dot{c}_j(x,t) \qquad (12)$$
$$+ \sum_{A=1}^{r} l_{IA}(x)\sum_{\rho=1}^{M}\partial_{j\mu_1\cdots\mu_\rho}h_A(\nabla^M c(t)(x))\dot{c}_{j,\mu_1\cdots\mu_\rho}(x,t).$$

Using (8) and (3), (12) may be expressed as

$$\dot{f}_I(x,t) = \int_B g_{Ii}(x,x')\int_B w_{ij}(x'-y)\partial_J p_j(\nabla^1 f(t)(y))F_J(f(t))(y)\,dx'dy$$
$$+ \int_B g_{Ii}(x,x')\int_B w_{ij}(x'-y)\partial_{J\mu_1}p_j(\nabla^1 f(t)(y))\{F_J(f(t))\}_{,\mu_1}(y)\,dx'dy$$
$$+ \sum_{A=1}^{r} l_{IA}(x)\partial_i h_A(\nabla^M c(t)(x))$$
$$\left[\begin{array}{l}\int_B w_{ij}(x-y)\partial_J p_j(\nabla^1 f(t)(y))F_J(f(t))(y)\,dy \\ + \int_B w_{ij}(x-y)\partial_{J\mu_1}p_j(\nabla^1 f(t)(y))\{F_J(f(t))\}_{,\mu_1}(y)\,dy\end{array}\right]$$
$$+ \sum_{A=1}^{r} l_{IA}(x)\sum_{\rho=1}^{M}\partial_{i\mu_1\cdots\mu_\rho}h_A(\nabla^M c(t)(x))$$
$$\left[\begin{array}{l}\int_B \dfrac{\partial^\rho w_{ij}}{\partial x_{\mu_1}\cdots\partial x_{\mu_\rho}}(x-y)\partial_J p_j(\nabla^1 f(t)(y))F_J(f(t))(y)\,dy \\ + \int_B \dfrac{\partial^\rho w_{ij}}{\partial x_{\mu_1}\cdots\partial x_{\mu_\rho}}(x-y)\partial_{J\mu_1}p_j(\nabla^1 f(t)(y))\{F_J(f(t))\}_{,\mu_1}(y)\,dy\end{array}\right]. \qquad (13)$$

Noting (1) and (11), the coarse fields and the gross determining functions are now seen to satisfy

$$F_I(\Gamma(\tilde{g},g,l,c(t)))(x) =$$
$$\int_B g_{Ii}(x,x')\int_B w_{ij}(x'-y)\partial_J p_j(\nabla^1\Gamma(\tilde{g},g,l,c(t))(y))F_J(\Gamma(\tilde{g},g,l,c(t)))(y)\,dx'dy$$
$$+ \int_B g_{Ii}(x,x')\int_B w_{ij}(x'-y)\partial_{J\mu_1}p_j(\nabla^1\Gamma(\tilde{g},g,l,c(t))(y))\{F_J(\Gamma(\tilde{g},g,l,c(t)))\}_{,\mu_1}(y)\,dx'dy$$
$$+ \sum_{A=1}^{r} l_{IA}(x)\partial_i h_A(\nabla^M c(t)(x))$$
$$\left[\begin{array}{l}\int_B w_{ij}(x-y)\partial_J p_j(\nabla^1\Gamma(\tilde{g},g,l,c(t))(y))F_J(\Gamma(\tilde{g},g,l,c(t)))(y)\,dy \\ + \int_B w_{ij}(x-y)\partial_{J\mu_1}p_j(\nabla^1\Gamma(\tilde{g},g,l,c(t))(y))\{F_J(\Gamma(\tilde{g},g,l,c(t)))\}_{,\mu_1}(y)\,dy\end{array}\right]$$
$$+ \sum_{A=1}^{r} l_{IA}(x)\sum_{\rho=1}^{M}\partial_{i\mu_1\cdots\mu_\rho}h_A(\nabla^M c(t)(x))$$

$$\left[\begin{array}{l}\int_B \dfrac{\partial^\rho w_{ij}}{\partial x_{\mu_1}\cdots\partial x_{\mu_\rho}}(\boldsymbol{x}-\boldsymbol{y})\partial_J p_j\left(\nabla^1\Gamma\left(\tilde{g},g,l,c(t)\right)(\boldsymbol{y})\right)F_J\left(\Gamma\left(\tilde{g},g,l,c(t)\right)\right)(\boldsymbol{y})d\boldsymbol{y}\\ +\int_B \dfrac{\partial^\rho w_{ij}}{\partial x_{\mu_1}\cdots\partial x_{\mu_\rho}}(\boldsymbol{x}-\boldsymbol{y})\partial_{J\mu_1}p_j\left(\nabla^1\Gamma\left(\tilde{g},g,l,c(t)\right)(\boldsymbol{y})\right)\left\{F_J\left(\Gamma\left(\tilde{g},g,l,c(t)\right)\right)\right\}_{,\mu_1}(\boldsymbol{y})d\boldsymbol{y}\end{array}\right]. \quad (14)$$

Motivated by the structure of (14), we now demand that the functions $\tilde{g}, g, l$ be such that

$$\text{(14), with } c(t) \text{ replaced everywhere by } c, \text{ and} \quad (15)$$

$$c_i(\boldsymbol{x}) = \int_B w_{ij}(\boldsymbol{x}-\boldsymbol{x}')p_j\left(\nabla^1\Gamma(\tilde{g},g,l,c)(\boldsymbol{x}')\right)d\boldsymbol{x}' \quad (16)$$

hold for **all** coarse fields $c$ and at all points $\boldsymbol{x}$ of the body.

In terms of a solution $\tilde{g}, g, l$ of (15) and (16), the evolutionary field equations of the ***coarse theory*** are defined by

$$\dot{c}_i(\boldsymbol{x},t) = \int_B w_{ij}(\boldsymbol{x}-\boldsymbol{y})\partial_J p_j\left(\nabla^1\Gamma(\tilde{g},g,l,c(t))(\boldsymbol{y})\right)F_J\left(\Gamma(\tilde{g},g,l,c(t))\right)(\boldsymbol{y})d\boldsymbol{y}$$
$$+\int_B w_{ij}(\boldsymbol{x}-\boldsymbol{y})\partial_{J\mu_1}p_j\left(\nabla^1\Gamma(\tilde{g},g,l,c(t))(\boldsymbol{y})\right)\left\{F_J\left(\Gamma(\tilde{g},g,l,c(t))\right)\right\}_{,\mu_1}(\boldsymbol{y})d\boldsymbol{y} \quad (17)$$
$$c_i(\boldsymbol{x},0) = c_{0i}(\boldsymbol{x}).$$

It is easy to see that each solution of (17) defines a solution of the fine theory (1). In terms of a solution $c$ of (17) generated using a particular solution of (15) and (16), define the functions $f$ of space and time according to (11). The time derivative of each $f_I$ so defined is given by the right hand side of (14). However, since $\tilde{g}, g, l$ are such that (15) is satisfied for all sets of coarse functions $c$, this implies that

$$\dot{f}_I(\boldsymbol{x},t) = F_I\left(\Gamma(\tilde{g},g,l,c(t))\right)(\boldsymbol{x}) = F_I(f(t))(\boldsymbol{x})$$
$$f_I(\boldsymbol{x},0) = \Gamma_I(\tilde{g},g,l,c_0). \quad (18)$$

The crucial factors for the above program to succeed, then, are the existence of an invariant manifold for the fine theory, a sufficiently general choice of the coarse variables so that it is capable of parametrizing the invariant manifold, and enough freedom in the chosen form of the Gross Determiner so that the invariant manifold can indeed be computed, i.e. (15) and (16) have at least one solution. There are no mathematical guarantees known to the author on any of these questions for the general class of theories we are interested in, so insight into the physics that the fine and coarse scale theories represent, in order to guess at a sufficient set of coarse variables, along with a fair amount of scientific optimism appear to be essential in exercising the method for practical purposes.

It is also clear that the above argument goes through just as well for chosen forms of the Gross determiner given by (6) or (7) with finitely many terms in the expansion. Also, while the procedure delivers complete field equations and constitutive equations for the coarse theory and indicates to the necessity of initial conditions, it does not deliver information on the nature of boundary conditions for the theory to be well-set. As for invariance requirements, they can be incorporated into the coarse scale theory as additional constraints to be satisfied by the Gross determiner, as discussed in Muncaster (1983; a).

In order to seek approximate solutions to (15) and (16), two finite sets of basis functions, for the coarse fields and the fine fields respectively, may be introduced. In terms of these basis functions the functions $\tilde{g}, g, l$ and $c$ may be discretized. By substituting these discrete functions in place of their exact counterparts, a least squares objective function can be generated from suitably non-dimensionalized versions of the equations in (15) and (16). This least squares residual can then be integrated over the body $\boldsymbol{B}$ as well as the parameter space represented by physically meaningful ranges of the coefficients for the discretization of the functions $c$ in terms of the coarse basis. The final objective function is to be minimized with respect to the coefficients of the discretization corresponding to the functions $\tilde{g}, g, l$, and acceptable solutions are those for which the minimum is zero.

This is a standard problem in nonlinear programming. A unique solution is not required – any solution suffices to define a coarse theory that is justified by the fact that its solutions, in turn, define a class of solutions to the fine theory. Of course, in case there exists more than one solution, the best candidate for the coarse theory may well be resolved by testing solutions of the corresponding coarse theory against experimental results. A scheme for numerical approximation based on the above ideas as well as the development of selection criteria for the 'optimal' coarse theory will be subjects of future research.

### 3. Model Problem: 1-D analysis of linearly rate-dependent elastic plastic material

We would like to examine the coarse response of a materially inhomogeneous, elastic-plastic, 1-D bar. The governing equations of the fine theory are taken to be

$$\left. \begin{aligned} \dot{u}(x,t) &= v(x,t) \\ \dot{v}(x,t) &= \frac{1}{\rho}\left[E\{u_x - p\}\right]_x(x,t) \\ \dot{p}(x,t) &= \omega(x)p(x,t) + \xi(x)u_x(x,t) \end{aligned} \right\} \quad 0 < x < 1. \tag{19}$$

In the above, a Latin subscript represents partial differentiation with respect to the subscript and a superposed dot represents a partial time derivative. The interval $[0,1]$ will be represented as $V$. $u$ represents the displacement field, $v$ is the velocity, $\rho$ is the density assumed to be uniform for simplicity, $E$ the spatially varying Young's modulus field, $p$ is the plastic strain field, and $\omega$ and $\xi$ are known inhomogeneous fields that define the linear plastic response. The latter two functions may be chosen appropriately to represent a linear dependence of the plastic strain rate on the stress. The linearity in the above system has been chosen intentionally in order to illustrate the theory in a simple context.

The intention now is to set up a coarse theory defined by ordinary differential equations whose solutions define a small, but hopefully useful, class of solutions to (19). To this end we define the average strain as the only coarse variable:

$$\bar{e}(t) = \int_V u_x(x,t)\,dx. \tag{20}$$

For the gross determining relationships, we choose the extremely restricted and simplified form

$$\begin{aligned}
u(x,t) &= \varphi(x)\bar{e}(t) \\
v(x,t) &= \psi(x)\bar{e}(t) \\
p(x,t) &= \mu(x)\bar{e}(t),
\end{aligned} \tag{21}$$

and the first task now is to determine the functions $\varphi$, $\psi$, and $\mu$ following the recipe set forth in Section 2. We express the LHS of (19) and (20) in terms of the functions $\varphi$, $\psi$, $\mu$, and $\bar{e}(t)$ and equate with the RHS of these equations, now expressed in terms of $\varphi$, $\psi$, $\mu$, and $\bar{e}(t)$ by substituting (21). The result of these operations is the following set of time independent equations, when all occurrences of $\bar{e}(t)$ are replaced by $\bar{e}$:

$$\begin{aligned}
\left[ \varphi(x) \int_V \psi_x(x)\,dx - \psi(x) \right] \bar{e} &= 0 \\
\left[ \psi(x) \int_V \psi_x(x)\,dx - \rho^{-1}\left\{ E(\varphi_x - \mu) \right\}_x(x) \right] \bar{e} &= 0 \\
\left[ \mu(x) \int_V \psi_x(x)\,dx - \{\omega\mu + \xi\varphi_x\}(x) \right] \bar{e} &= 0 \\
\left[ 1 - \int_V \varphi_x(x)\,dx \right] \bar{e} &= 0.
\end{aligned} \tag{22}$$

We now demand that (22) hold for all possible states of the coarse theory, i.e. for all values of the number $\bar{e}$. As a consequence, the governing equations for the functions $\varphi$, $\psi$, and $\mu$ become

$$\varphi(x)[\psi(1)-\psi(0)]=\psi(x)$$
$$\psi(x)[\psi(1)-\psi(0)]=\rho^{-1}\{E(\varphi_x-\mu)\}_x(x)$$
$$\mu(x)[\psi(1)-\psi(0)]=\{\omega\mu+\xi\varphi_x\}(x) \tag{23}$$
$$\varphi(1)-\varphi(0)=1.$$

A necessary condition for the satisfaction of (23) is

$$\left\{\frac{1}{\rho}E\left[1-\frac{\xi}{[\psi(1)-\psi(0)-\omega]}\right]\varphi_x\right\}_x-[\psi(1)-\psi(0)]^2\varphi=0 \tag{24}$$
$$\varphi(1)-\varphi(0)=1.$$

We now fix the value of
$$\psi(1)-\psi(0)=S>0, \tag{25}$$

and assume that the prescribed data is such that
$$1-\frac{\xi(x)}{[S-\omega(x)]}>0. \tag{26}$$

As demonstrated at the end of this section, these assumptions are valid ones to have a coarse theory corresponding to a fine creep law of the form $\dot{p}=C\sigma, C>0$ such that the tangent modulus of the coarse stress-strain curve is positive and less than the elastic modulus. (24) is a linear, second-order 2-point boundary value problem for $\varphi$. We assume that by fixing $\varphi(0)$ and $S$ we are able to obtain a solution to (24). In terms of such a solution, we define

$$\psi(x):=S\varphi(x)$$
$$\mu(x):=\frac{\xi(x)\varphi_x(x)}{\{S-\omega(x)\}}; \tag{27}$$

because $\varphi$ is such that $\varphi(1)-\varphi(0)=1$, a solution to (23) has been defined by the solution of (24) and the functions defined in (27). Solutions to (23) are clearly non-unique, a family parametrized by the values of $S$ and $\varphi(0)$ having been derived. It may also be checked that for the chosen form of the gross determining relationship given by (21), the procedure does not succeed in defining a coarse scale inelastic theory if $\xi=0, \omega\neq 0,$, i.e. if a coupling of the plastic strain evolution with the total strain is absent.

For fixed $S$, the corresponding coarse scale theory is given by the ordinary differential equation

$$\dot{\bar{e}}(t)=\bar{e}(t)\int_V \psi_x(x)dx=[\psi(1)-\psi(0)]\bar{e}(t) \tag{28}$$

and hence

$$\dot{\bar{e}}(t) = S\bar{e}(t)$$
$$\bar{e}(0) = \bar{e}_0, \tag{29}$$

with the obvious solution

$$\bar{e}(t) = \bar{e}_0 \exp(St). \tag{30}$$

The class of solutions of the fine scale theory (19), defined by the derived coarse scale theory (29), is parametrized by the quantities $\bar{e}_0, S, \varphi(0)$ and may be expressed as

$$u(x,t;\bar{e}_0,S,\varphi(0)) = \varphi(x;S,\varphi(0))\bar{e}_0 \exp(St),$$
$$v(x,t;\bar{e}_0,S,\varphi(0)) = S\varphi(x;S,\varphi(0))\bar{e}_0 \exp(St), \tag{31}$$
$$p(x,t;\bar{e}_0,S,\varphi(0)) = \frac{\xi(x)\varphi_x(x;S,\varphi(0))}{[S-\omega(x)]}\bar{e}_0 \exp(St).$$

As a consequence of the definition of the variable $\bar{e}(t)$, this class of solutions to the fine scale theory may be interpreted as representing the situation where the left end of the 1-D bar at $x=0$ is fixed and the end at $x=1$ is pulled out or pushed in according to

$$\dot{u}(1,t) = Su(1,t)$$
$$u(1,0) = L\bar{e}_0, \tag{32}$$

where $L$ is a constant of unit magnitude with dimension of length.

The class of solutions (31) is indeed restricted; for instance non-trivial solutions of (19) beginning from an unstrained state do not belong to it (however, arbitrarily small strained initial states may be considered), and only exponentially growing extensional and compressional strain solutions may be considered. We also note that if $\xi(x) = E(x)C(x)$, $C > 0$, and $\omega(x) = -\xi(x)$ so that $\dot{p}(x,t) = C(x)\sigma(x,t)$ in the fine theory, where $\sigma = E(u_x - p)$ represents the stress, then for the choice $S = 0$ representing a relaxation test, the fine solution that is recovered is that corresponding to beginning from a fully relaxed state. But this is to be expected – the states of the fine theory accessible through states of the coarse theory is necessarily limited as, roughly speaking, the 'degrees of freedom' of the coarse theory are much less than the fine theory. For instance, since the only degree of freedom is the average strain, which is fixed in a relaxation test, it is clear that to recover details of stress evolution in a relaxation test greater generality is required in the choice of the gross determiner (21). In a way, what is remarkable about the procedure is that, even under such severe constraints, it does not produce an inconsistent physical result under circumstances corresponding to the relaxation test.

If we define the stress and plastic deformation of the coarse theory as

$$\bar{\sigma}(t) = \int_V \sigma(x,t) dx \quad ; \quad \bar{p}(t) = \int_V p(x,t) dx, \tag{33}$$

then

$$\bar{\sigma}(t) = \bar{e}(t) \int_V E(x) \varphi_x(x;S) \left\{ 1 - \frac{\xi(x)}{[S - \omega(x)]} \right\} dx,$$

$$\bar{p}(t) = \bar{e}(t) \int_V \frac{\xi(x) \varphi_x(x;S)}{[S - \omega(x)]} dx. \tag{34}$$

In the case when $E, \xi, \omega$ are uniform and $\xi = -\omega$ with $\omega < 0$ (as would be required to have $\dot{p} = C\sigma, C > 0$), a stress-strain relationship may be explicitly written as

$$\bar{\sigma}(t) = \left\{ E\left(\frac{S}{S - \omega}\right) \right\} \bar{e}(t). \tag{35}$$

Solutions of the coarse theory in such instances reflect a strain-rate hardening effect at fixed strain for increasing $S > 0$, with the tangent modulus $\partial \bar{\sigma}/\partial \bar{e} < E$ and approaching $E$ as $S \to \infty$ corresponding to instantaneous deformation.

## 4. Model Problem: Reduction of fields in idealized continuum dislocation mechanics

The concern here is to derive a coarse theory that is not a spatial homogenization of a fine theory but one that strictly reduces the number of variables of the fine theory.

For the fine theory, we assume a continuum with only straight edge dislocations with burgers vector in the $y$ direction and with line direction along the $z$ axis. These dislocations can move only in the $y$ direction, i.e. single slip with the slip plane being the $y - z$ plane with normal in the $x$ direction. Let $\omega$ be the number density of such dislocations per unit area perpendicular to the $z$ axis. We assume that all fine fields vary only with respect to $x$ and $y$. Then, it can be shown (Acharya 2001; a, b) that the evolution of the field $\omega$ is governed by the equation

$$\dot{\omega}(x,y,t) = -\{V(\sigma,\omega)\omega\}_y \tag{36}$$

in the absence of dislocation nucleation, where $V$ is the magnitude of the dislocation velocity field that will depend on the arguments shown as well as the magnitude of the Burgers vector, $b$, and a dislocation drag coefficient, $B$ (physical dimension of stress.time). In the above $\sigma$ represents the stress component that characterizes the shear traction in the $y$ direction on the plane with normal in the $x$ direction. Motivated by the form of the driving force for dislocation velocity that may be derived from thermodynamic considerations (Acharya 2001; b), we assume a velocity following a linear drag relationship

$$V = \frac{\sigma \omega b C b^2}{B}, \qquad (37)$$

where the numerator represents the force per unit length on the dislocations threading a cross sectional area of $Cb^2$, where $C$ is a nondimensional constant. With these assumptions, the equations of the fine theory may be written as

$$\dot{\omega} = -\left(\frac{Cb^3}{B}\sigma\omega^2\right)_y$$

$$\dot{p} = A\sigma\omega^2 \quad ; \quad A = \left(\frac{Cb^4}{B}\right) \qquad (38)$$

where $p$ is the $yx$ component of the plastic shear distortion. The above set of equations is to be supplemented with the vectorial total displacement ($\mathbf{u}$)- velocity ($\mathbf{v}$) relationships and balance of linear momentum in which the stress tensor is a function of the displacement gradient and the plastic distortion tensors. We do not write these out as they do not play a role in the subsequent argument. In the present case, the only non-zero component of the plastic distortion tensor is $p$.

We choose the variables of the coarse theory to be all the fields of the fine theory *except* $\omega$, i.e.

$$\bar{\mathbf{u}} = \mathbf{u}$$
$$\bar{\mathbf{v}} = \mathbf{v} \qquad (39)$$
$$\bar{p} = p$$

which also serve as the gross determining relationships for the corresponding fine fields. Based on (39), $\bar{\sigma} = \sigma$. We assume the gross determiner for $\omega$ to be of the form

$$\omega(x,y,t) = a_1 p(x,y,t) + a_2 p_y(x,y,t) + a_3 p_{yy}(x,y,t) =: \hat{\omega}(a,p). \qquad (40)$$

The only non-trivial equation that has to be solved to define our coarse scale theory is the following:

$$-\left\{\frac{A}{b}\sigma[\hat{\omega}(a,p)]^2\right\}_y = a_1\left\{A\sigma[\hat{\omega}(a,p)]^2\right\} + a_2\left\{A\sigma[\hat{\omega}(a,p)]^2\right\}_y + a_3\left\{A\sigma[\hat{\omega}(a,p)]^2\right\}_{yy}, \qquad (41)$$

which has to hold for all possible fields $\sigma, p$. Clearly, one solution of (41) is given by

$$a_1 = 0, \, a_2 = -1/b, \, a_3 = 0. \qquad (42)$$

The coarse theory, in terms of this solution is given by

$$\dot{p} = \frac{Cb^2}{B}\sigma(p_y)^2, \qquad (43)$$

along with the vectorial equation of balance of linear momentum for an elastic plastic material and the displacement velocity relationship. The dislocation density variable has been eliminated in the coarse theory resulting, interestingly, in a first gradient-dependent, gradient plasticity model. We emphasize that each solution of the derived gradient plasticity theory defines a solution of the fine theory representing dislocation mechanics.

## 5. Model Problem: Formal structure of homogenized, 1-D elastodynamics

Here we provide a formal derivation of a coarse theory for the response of a materially inhomogeneous, elastic, 1-D bar. Unlike the model problem considered in Section 3, here we would like to have coarse a theory that admits the imposition of boundary conditions.

The governing equations of the fine theory are taken to be

$$\left. \begin{array}{l} \dot{u}(x,t) = v(x,t) \\ \dot{v}(x,t) = \dfrac{1}{\rho}\left[Eu'\right]'(x,t) \end{array} \right\} \quad 0 < x < 1. \tag{44}$$

The interval $[0,1]$ will be represented as $V$. $u$ represents the displacement field, $v$ is the velocity, $\rho$ is the density assumed to be uniform for simplicity, and $E$ the spatially varying Young's modulus field. In this section, we shall use a prime to denote the partial derivative with respect to the space variable.

We choose the coarse variables to be running spatial averages, over specified length scales, of the displacement and velocity fields:

$$\begin{aligned} \bar{u}(x,t) &= \int_V w(x,y)u(y,t)dy \\ \bar{v}(x,t) &= \int_V w(x,y)v(y,t)dy, \end{aligned} \tag{45}$$

where the known weighting function $w$ sets the length scale of averaging. Next we assume the gross determining relationship to be of the form

$$\begin{aligned} u(x,t) &= \int_V \varphi(x,y)\bar{u}(y,t)dy \\ v(x,t) &= \int_V \psi(x,y)\bar{v}(y,t)dy, \end{aligned} \tag{46}$$

and the primary task is to determine the equations governing the kernels $\varphi$ and $\psi$.

Following the recipe set forth in Section 2, we utilize the fine scale equations and express them completely in terms of the coarse fields and the gross determining functions. From the fine scale displacement velocity relationship we obtain

$$\begin{aligned}
\dot{u}(x,t) &= \int_V \varphi(x,y)\dot{\bar{u}}(y,t)dy \\
&= \int_V \varphi(x,y)\int_V w(y-z)\dot{u}(z,t)dzdy = \int_V \varphi(x,y)\int_V w(y-z)v(z,t)dzdy \\
&= \int_V \varphi(x,y)\int_V w(y-z)\int_V \psi(z,r)\bar{v}(r,t)drdzdy \\
&= v(x,t) = \int_V \psi(x,r)\bar{v}(r,t)dr.
\end{aligned} \quad (47)$$

Consequently, we require that

$$\int_V \left[\int_V \varphi(x,y)\int_V w(y-z)\psi(z,r)dzdy - \psi(x,r)\right]\bar{v}(r)dr = 0, \quad (48)$$

for all fields $\bar{v}$, a condition that is satisfied if

$$\int_V \varphi(x,y)\int_V w(y-z)\psi(z,r)dzdy = \psi(x,r), \quad (49)$$

which is an integral equation for the kernels $\varphi$ and $\psi$.

Similarly, from the fine scale balance of linear momentum equation we obtain

$$\begin{aligned}
\dot{v}(x,t) &= \int_V \psi(x,y)\dot{\bar{v}}(y,t)dy = \int_V \psi(x,y)\int_V w(y-z)\dot{v}(z,t)dzdy \\
&= \int_V \psi(x,y)\int_V w(y-z)\rho^{-1}(Eu')'(z)dzdy \\
&= \int_V \psi(x,y)\int_V w(y-z)\rho^{-1}\left[E(\cdot)\left\{\int_V \varphi(\cdot,r)\bar{u}(r,t)dr\right\}'\right]'(z)dzdy \\
&= \rho^{-1}(Eu')'(x,t) = \rho^{-1}\left[E(\cdot)\left\{\int_V \varphi(\cdot,r)\bar{u}(r,t)dr\right\}'\right]'(x).
\end{aligned} \quad (50)$$

Denoting by $\partial_1\varphi$ and $\partial_{11}\varphi$ the first and second partial derivative functions of $\varphi$ with respect to its first argument, (50) indicates that we need to require that

$$\int_V \left[\begin{array}{c}\int_V \psi(x,y)\int_V w(y-z)\{E'(z)\partial_1\varphi(z,r)+E(z)\partial_{11}\varphi(z,r)\}dzdy \\ -\{E'(x)\partial_1\varphi(x,r)+E(x)\partial_{11}\varphi(x,r)\}\end{array}\right]\bar{u}(r)dr = 0 \quad (51)$$

holds for all functions $\bar{u}$. This is satisfied if

$$\begin{aligned}
\int_V \psi(x,y)\int_V w(y-z)&\{E'(z)\partial_1\varphi(z,r)+E(z)\partial_{11}\varphi(z,r)\}dzdy \\
&= \{E'(x)\partial_1\varphi(x,r)+E(x)\partial_{11}\varphi(x,r)\},
\end{aligned} \quad (52)$$

which is another integro-differential constraint to be satisfied by the kernels $\varphi$ and $\psi$.

In addition, the following conditions also have to be satisfied by the kernels:

$$\begin{aligned}
\bar{u}(x) &= \int_V w(x-y)\int_V \varphi(y,z)\bar{u}(z)dzdy \text{ for all } \bar{u} \\
\bar{v}(x) &= \int_V w(x-y)\int_V \psi(y,z)\bar{v}(z)dzdy \text{ for all } \bar{v}.
\end{aligned} \quad (53)$$

Equations (49), (52), and (53) form the set of equations for determining the gross determining relationship. In the event that a solution *exists* to this set of equations, the corresponding coarse theory may be expressed as

$$\dot{\bar{u}}(x,t) = \bar{v}(x,t)$$

$$\dot{\bar{v}}(x,t) = \int_V w(x-z)\rho^{-1}\left[E(\cdot)\left\{\int_V \varphi(\cdot,r)\bar{u}(r,t)dr\right\}'\right](z)dz. \qquad (54)$$

Denoting

$$\sigma_G(x,t) = E(x)\left\{\int_V \partial_1\varphi(x,r)\bar{u}(r,t)dr\right\}$$

$$\bar{\sigma}(x,t) = \int_V w(x-z)\sigma_G(z,t)dz, \qquad (55)$$

the second of (54) may be alternatively expressed as

$$\rho\ddot{\bar{u}}(x,t) = \partial_x\bar{\sigma}(x,t) + w(x-1)\sigma_G(1,t) - w(x)\sigma_G(0,t). \qquad (56)$$

A corresponding formal derivation may be given for generating the coarse equations of balance of linear momentum for a gradient approximation of the gross determining relationship. That derivation would make it clear that boundary conditions can indeed be admitted.

In case (49), (52), and (53) have a solution, (56) may be used to study wave propagation in the coarse displacement solution, including possible dispersion and attenuation effects.

## 6. *Some observations on the range of applicability of the method*

Interestingly, it is clear that the formal procedure presented herein also applies to developing a field theory of the average response of a collection of particles interacting according to Newton's laws for a specified inter-particle force law (system of nonlinear ODEs), or a system of bodies whose dynamics is governed by Newton's laws and forces arising from inter-particle, elastic-viscous collisions. The key questions for the success of the homogenization of these fine scale theories according to the proposed procedure are whether invariant manifolds exist for them, and if so, the appropriate level of generality to be adopted in the definition of the Gross Determining relationship to obtain practically computable coarse theories. As it is known for some finite dimensional dynamical systems, e. g. the 3-degree-of-freedom Lorenz model for thermal convection, while invariant manifolds may exist they may not be amenable to a continuous parametrization[i], in which case more sophisticated mathematical tools than used here would be required to execute the main idea of the method of invariant manifolds.

---

[i] I am grateful to Shlomo Ta'asan for a discussion of this matter.

In the form the method has been developed in this paper, the coarse theory may only be interpreted as governing equations for coarse fields related to sufficiently smooth solutions of the fine theory. However, it is well known that most fine theories admit physically relevant non-smooth solutions. The governing equations for coarse fields pertaining to such solutions can perhaps be generated by the same conceptual ideas but much more sophisticated tools from mathematical analysis. As a practical device, and in the spirit of the development of constitutive *assumptions* in macroscopic theories, the governing equations for the coarse fields corresponding to smooth fine fields may simply be declared to be valid for coarse fields corresponding to non-smooth solutions as well. In looking for whether a rational basis for such a claim can exist at all, a seemingly interesting mathematical question arises in the homogenization context: given a formula for generating coarse fields from fine fields, does there always exist a smooth solution, say A, to the fine theory corresponding to an arbitrary solution of it, say B, such that the coarse fields corresponding to both A and B are identical?

## 7. *References*

Acharya, A. (2001 a) A model of crystal plasticity based on the theory of continuously distributed dislocations, *Journal of the Mechanics and Physics of Solids,* **49,** 761-784.

Acharya A. (2001 b) Driving forces and boundary conditions in continuum dislocation mechanics, submitted.

Amirat, Y., Hamdache, K., and Ziani, A. (1992) Homogenisation of parametrised families of hyperbolic problems, *Proceedings of the Royal Society of Edinburgh,* **120A**, 199-221.

Carr, J. and Muncaster, R. G. (1983 a) The application of centre manifolds to amplitude expansions. I. Ordinary differential equations, *Journal of Differential Equations,* **50**, 260-279.

Carr, J. and Muncaster, R. G. (1983 b) The application of centre manifolds to amplitude expansions. II. Infinite dimensional problems, *Journal of Differential Equations,* **50**, 280-288.

Cohen, H. and Muncaster, R. G. (1984) The dynamics of pseudo-rigid bodies: general structure and exact solutions, *Journal of Elasticity,* **14**, 127-154.

DiPerna, R. J. (1985) Measure-valued solutions to conservation laws, *Archive for Rational Mechanics and Analysis,* **88**, 223-270.

Kocks, U. F., Tomé, C. N., and Wenk, H.-R. (1998) Texture and anisotropy, Cambridge University Press, Cambridge.

McLaughlin, D., Papanicolaou, G., and Tartar, L. (1985) Weak limits of semilinear hyperbolic systems with oscillating data, *Lecture Notes in Physics,* **230**, Springer Verlag, 277-289.

Muncaster, R. G. (1983 a) Invariant manifolds in mechanics I: The general construction of coarse theories from fine theories, *Archive for Rational Mechanics and Analysis*, **84**, 353-373.

Muncaster, R. G. (1983 b) Invariant manifolds in mechanics II: Zero-dimensional elastic bodies with directors, *Archive for Rational Mechanics and Analysis*, **84**, 375-392.

Sabina, F. J., Smyshlyaev, V. P., and Willis, J. R. (1993) Self-consistent analysis of waves in a matrix-inclusion composite – I. Aligned spheroidal inclusions, *Journal of the Mechanics and Physics of Solids,* **41**, 1573-1588.

Smyshlyaev, V. P., Willis, J. R., and Sabina, F. J. (1993 a) Self-consistent analysis of waves in a matrix-inclusion composite – II. Randomly oriented spheroidal inclusions, *Journal of the Mechanics and Physics of Solids,* **41**, 1589-1598.

Smyshlyaev, V. P., Willis, J. R., and Sabina, F. J. (1993 b) Self-consistent analysis of waves in a matrix-inclusion composite – II. A matrix containing cracks, *Journal of the Mechanics and Physics of Solids,* **41**, 1809-1824.

Tartar, L. (1990) H-measures, a new approach for studying homogenization, oscillations and concentration effects in partial differential equations, *Proceedings of the Royal Society of Edinburgh,* **115A**, 193-230.

Truesdell, C. A. and Muncaster, R. G. (1980) Fundamentals of Maxwell's kinetic theory of a simple monatomic gas, Academic Press, New York.

E, W. and Kohn, R. V. (1991) The initial-value problem for measure-valued solutions of a canonical 2x2 system with linearly degenerate fields, *Communications on Pure and Applied Mathematics*, **XLIV**, 981-1000.

Willis, J. R. (1981) Variational and related methods for the overall properties of composites, In: *Advances in Applied Mechanics,* **21**, 2-77.